\documentclass[]{revtex4}
\usepackage{hyperref}
 
\begin{document}

\title{Gaugeon Formalism for Perturbative Quantum Gravity }

\author{ Sudhaker Upadhyay}
 \email {  sudhakerupadhyay@gmail.com; 
 sudhaker@boson.bose.res.in}

\affiliation { S. N. Bose National Centre for Basic Sciences,\\
Block JD, Sector III, Salt Lake, Kolkata -700098, India. }
 
\begin{abstract}
In this paper we investigate the Yokoyama  gaugeon formalism for perturbative quantum gravity
in a general curved spacetime.
Within the gaugeon formalism, we extend the configuration space   by introducing
vector gaugeon fields describing a quantum gauge degree of freedom. Such an extended  theory of perturbative gravity admits quantum gauge transformations leading to a  natural shift in the gauge parameter. 
Further we impose the Gupta-Bleuler type subsidiary condition to remove the
unphysical gaugeon modes. To replace the Gupta-Bleuler type condition by
more acceptable Kugo-Ojima type subsidiary condition we analyze the BRST symmetric 
gaugeon formalism.  Further,  the physical
Hilbert space is constructed for the perturbative quantum gravity which remains
  invariant under both the BRST symmetry and the quantum gauge transformations.
 
  \end{abstract}
\maketitle

\section{  Introduction}
The usual perturbative approach of covariant quantum gravity   in curved spacetime starts with the
Einstein-Hilbert theory and expands the full Riemannian metric   around a constant background.
The diffeomorphism invariance then translates into a gauge symmetry of the fluctuation  \cite{sw}.
Consequently, the problem of formulating the corresponding quantum field theory  in
general curved spacetime
 is conceptually not much different from Yang-Mills theory. 
The study of quantum field theory, particularly, in de Sitter spacetime  is very important as
it gets  relevance in inflationary cosmologies \cite{haw, hg, adl, aa}. 
Recent observations indicate that our universe is expanding in such a
rate that it may approach de Sitter spacetime asymptotically \cite{per}. 
By the gauge invariant  perturbative quantum gravity in curved space one
has attempted in a great effort  to unify gravity with Maxwell theory \cite{ein}. The
gauge invariant gravity models  have their relevance in string theories also \cite{ch,da,ah}.
Recently, significant developments have been made in the subject of  quantum gravity  in 
various directions \cite{Ana, mirh,eyo,faiza,mf1,mf2,mf3,mf4,mf5,mf6,mf7,mf8,mf9,mf10,mir2,mf11,mf12,mf13,mf14,mf15,
mf16,mf17,mf18,mf19}.

On the other hand, the covariant quantization of perturbative gravity in general curved 
spacetime  cannot
be done without getting rid of the redundant degrees of freedom as the classical theory is gauge 
invariant \cite{hig}.  The spurious degrees of freedom in the theory of perturbative gravity are 
removed
by imposing a covariant gauge condition \cite{hig}.  
 The gauge  conditions are incorporated  
at quantum level of the theory by adding   suitable  gauge-fixing and ghost terms to the classical action, which 
remains invariant under the fermionic rigid  BRST  transformation \cite{faiza, upa, upa1}.
  However, in the standard   quantization of gauge theories, one does
not consider the gauge transformation at the quantum level as there is no quantum
gauge freedom. The quantum theory is defined only after fixing the gauge.
  Hayakawa and Yokoyama have shown that
  a shift in gauge parameter occurs through renormalization 
  which affects the gauge-fixing condition \cite{haya}.
  
Yokoyama's gaugeon formalism \cite{yo0,yok,yo1,yo2,yo3} provides a wider framework to quantize the 
gauge theories  in which we 
discuss the quantum gauge transformation. 
The shift of the
gauge parameter through renormalization has been naturally derived from the gauge structure within this
formalism \cite{yo0}. In this formalism, we extend  the configuration space by introducing
a set of extra fields (so-called gaugeon fields) in the effective Lagrangian density describing  the quantum gauge freedom.  It is obvious that the gaugeon modes do not contribute to physical processes and therefore one needs to remove them. First of all, Yokoyama   achieved this by 
putting the Gupta-Bleuler type constraint on the gaugeon field, which has its own limitation \cite{yo0}. Further, by introducing the
BRST symmetric  gaugeon formulation  this situation is improved \cite{ki,mk} which is
facilitated 
by a more acceptable Kugo-Ojima type restriction \cite{kugo, kugo1}. 
The gaugeon formalism has been studied many times for the gauge theories in flat space time
\cite{ki, mk, mk1, naka, rko, miu, mir1, mir2}; however, it has not been discussed  in the context of gauge theories in curved spacetime.
This provides a motivation   to extend  such  a 
formalism for the quantum theory of gravity  in curved
spacetime. We show that this formalism  also holds for the theory of linearized gravity where the 
fluctuations of the metric are treated as the gauge field. 

In this paper, we consider the theory of perturbative quantum gravity
in general curved metric space to discuss both the gauge and the BRST invariance. Further, to analyze the quantum gauge freedom  of the theory, we extend the effective action by introducing two vector gaugeon fields. Within the gaugeon formalism, we investigate the quantum gauge transformation under which such an extended  Lagrangian density (also called  
the Yakoyama Lagrangian density) remains form  invariant.
The transformed fields under a quantum gauge transformation   satisfy the same equations 
of motion as the original ones, but with a shifted gauge parameter. Furthermore, we 
implement two subsidiary conditions, of Kugo-Ojima type 
and Gupta-Bluler type,
to remove the unphysical graviton and gaugeon modes, respectively. 
After that we demonstrate  the BRST symmetric gaugeon formalism for 
perturbative gravity theory by further introducing  
ghost fields corresponding to the gaugeon fields. Such a BRST symmetric gaugeon action
possesses  both the BRST  symmetry
and the quantum gauge transformations.
Further we show that by virtue of the BRST symmetry both 
the Kugo-Ojima type and the Gupta-Bluler type  subsidiary conditions get converted into 
 a single Kugo-Ojima type condition. Finally, the physical Hilbert space is  constructed for
the quantum gravity, which is annihilated by the BRST charge and also remains  invariant under quantum gauge transformations. 

We organize the paper as follows.
In section II, we present the  perturbative quantum gravity in general curved spacetime having
gauge and BRST invariance. Section III is devoted to the study of the standard gaugeon formalism
for perturbative quantum gravity. In section IV, BRST symmetric gaugeon formalism is discussed. In the last section, we summarize our work. 
 
\section{ The perturbative quantum gravity in curved spacetime}
In this section, we  analyze  the BRST 
symmetry of  perturbative quantum gravity   in  general curved spacetime.
For this purpose,  we begin with
the Lagrangian density for the theory of classical gravity
in general  curved spacetime   defined by
\begin{equation}
{\cal L}_c   =   \sqrt{ -\tilde g}  (R- 2\Lambda ), \label{kin}
\end{equation}
where $\tilde g$, $R$ and $\Lambda$ are the determinant of the full metric $\tilde g_{ab}$, the Ricci scalar curvature, and the cosmological constant, respectively. Here  units  are adopted such that $16\pi G=1$. The Lagrangian density remains invariant under the following infinitesimal
transformation  originating from its general coordinate invariance:
\begin{eqnarray}
\delta_\rho \tilde g_{ab} =    \nabla_a \rho_b +  \nabla_b \rho_a,\label{gau}
\end{eqnarray}
where $\nabla_a$ denotes the background covariant derivative and $\rho_a$ represents a vector field.
In perturbative  quantum gravity, one writes the full metric in terms of a fixed
background metric and small perturbations around it.
Therefore, we  decompose the full metric $\tilde g_{ab} $ of classical gravity  as
\begin{equation}
\tilde g_{ab} =g_{ab}+h_{ab},
\end{equation}
where $g_{ab}$ refers to the  fixed background metric 
and $h_{ab}$ refers to  small perturbations around the fixed metric.
With the help of the above decomposition one can express the Lagrangian density for perturbative quantum gravity (\ref{kin}) in terms of the fluctuation $h_{ab}$. However, after being decomposed  the transformation of $\tilde g_{ab}$ mentioned in (\ref{gau})  
will be attributed to $h_{ab}$
as follows:
 \begin{eqnarray}
\delta_\rho  h_{ab} 
&=&\nabla_a \rho_b +\nabla_b \rho_a.
\end{eqnarray}
 The gauge invariance of the Lagrangian density (\ref{kin}) implies that there are redundancies in the physical 
 degrees  of freedom.  These redundancies of the degrees of 
freedom give rise to constraints in the canonical quantization  \cite{ht} and produce 
divergences in the generating functional   in the path integral  quantization. In 
order to remove these 
redundancies we need  to break the local gauge covariance   by  fixing the gauge as follows: \cite{faiza}
  \begin{equation}
G[h]_a=(\nabla^b h_{ab} -k\nabla_a h) =0,
\end{equation}
where $k\neq  1$ is the gauge parameter. For $k=1$ the conjugate momentum corresponding to $h_{00}$ 
vanishes
and therefore the partition function becomes ambiguous. 
To avoid such ambiguities sometimes $k$ is written in terms of an arbitrary finite constant $\beta$ as follows: $k=(1+\beta)/ \beta $ \cite{hig}.

To  incorporate the   above gauge-fixing condition in the theory of linearized gravity
 at the quantum level we add the following covariant gauge-fixing term in the 
 gauge invariant Lagrangian density of pure gravity:
 \begin{eqnarray}
{\cal L}_{gf}= \sqrt{- g}[ b^a(\nabla^b h_{ab}-k \nabla_a h) +\frac{\alpha}{2} b_ab^a], \label{gfix} 
\end{eqnarray} 
where $\alpha$ is a gauge parameter and $b^a$ is a Nakanishi-Lautrup type auxiliary field. 

Further, to compensate the contribution of the above gauge-fixing term in the functional integral we need to add the
following Faddeev-Popov ghost term   
in the effective theory:
  \begin{eqnarray}
{\cal L}_{gh}&=&   \sqrt{- g}\bar c^a \nabla^b [ \nabla_a c_b+ \nabla_b c_a- 
2kg_{ab}\nabla_c c^c ], \nonumber\\
  &=&\sqrt{ -g}\bar c^a M_{ab} c^b,
\end{eqnarray} 
where the Faddeev-Popov matrix operator ($M_{ab}$)  has the following form:
\begin{eqnarray} 
M_{ab} =  \nabla_c \left[ \delta_b^c\nabla_a  + g_{ab}\nabla^c - 2k \delta_a^c\nabla_b \right].
 \end{eqnarray}
Here, we note that  the Faddeev-Popov ghost ($c^a$) and anti-ghost
 ($\bar c^a$) fields appearing in the theory of perturbative gravity are vector fields.

Now, the total effective Lagrangian density 
for perturbative quantum gravity in covariant gauge  is defined by 
\begin{equation}
 {\cal L}_{T} = {\cal L}_c  +{\cal L}_{gf}+{\cal L}_{gh},  \label{com}
\end{equation}
which admits the following nilpotent BRST  transformation:
\begin{eqnarray}
&&s  h_{ab} =  - (\nabla_a c_b +\nabla_b c_a ), \nonumber\\ 
&&s c^a  = -c_b\nabla^b c^a,\nonumber\\
&& s  \bar c^a
= b^a, \ 
 s  b^a =  0. \label{sym}
\end{eqnarray}
We observe that the sum of the gauge-fixing and ghost parts of the effective Lagrangian density  (i.e.,
$ {\cal L}_{gf} +{\cal L}_{gh}=:{\cal L}_g$) 
 is BRST exact, and with the help of the above BRST symmetry it can be expressed  as  \cite{mf8}
\begin{eqnarray}
{\cal L}_g  
  &=&  s \sqrt{ -g}\left[\bar c ^a \left(\nabla^b h_{ab} -k\nabla_a h +\frac{\alpha}{2} b_a\right)\right],\nonumber\\
&=& s  \Psi,\label{g}
\end{eqnarray}
 where  $\Psi$ denotes the gauge-fixing fermion of the theory  
with the following expression:
\begin{equation}
\Psi =  \sqrt{ -g}\left[\bar c ^a \left(\nabla^b h_{ab} -k\nabla_a h +\frac{\alpha}{2} b_a\right)\right].\label{gff}
\end{equation}
In the next section, we will study the 
 development of the quantum gauge transformation for the
  covariant linearized gravity theory using the standard gaugeon formalism.
\section{Yokoyama gaugeon formalism }
In this section, we analyze the quantum gauge transformations
using the Yokoyama gaugeon formalism  for
perturbative quantum gravity in a general curved metric space. For this purpose, we construct
the Yokoyama Lagrangian density for perturbative quantum gravity
   by incorporating the vector gaugeon fields $y^a$ and $y^a_\star$   satisfying 
Bose-Einstein statistics as follows:
\begin{eqnarray}
{\cal L}_{yk}& =& {\cal L}_c +\sqrt{- g} b^a(\nabla^b h_{ab}-k \nabla_a h)  +\frac{\varepsilon}{2} \sqrt{- g}(
y^a_\star + \lambda b^a )^2 
+\sqrt{ -g}\bar c^a M_{ab} c^b \nonumber\\
&+&  \sqrt{- g} \nabla^b y_\star ^a [ \nabla_a y_b+ \nabla_b y_a- 
2kg_{ab}\nabla_c y^c ],\label{yk}
\end{eqnarray}
where  $\varepsilon$ is a sign factor $(=\pm 1)$
and $\lambda$ is the gauge parameter, which is identified with $\alpha$ of (\ref{gfix}) as $\alpha 
=\varepsilon\lambda^2$.

The Lagrangian density (\ref{yk}) admits the quantum gauge transformation which enables us to vary the gauge parameter.
 The quantum gauge transformation is given by
 \begin{eqnarray}
&& h_{ab} \rightarrow\hat h_{ab} =   h_{ab} -\tau( \nabla_a y_b +  \nabla_b y_a),\nonumber\\
&&y^a_\star\rightarrow \hat y^a_\star = y^a_\star - \tau b^a,\nonumber\\
&& y_a\rightarrow\hat y_a =y_a, 
\nonumber\\
&& b^a\rightarrow \hat b^a =b^a,\nonumber\\
&& \bar c^a\rightarrow \hat{\bar c}^a  =\bar c^a,\nonumber\\
&& c^a\rightarrow \hat c ^a =c^a,  
  \end{eqnarray}
where $\tau$ is an infinitesimal transformation parameter of a bosonic nature. Under such a 
quantum gauge transformation  the Lagrangian density (\ref{yk}) remains ``form invariant",
 i.e. it transforms as 
\begin{eqnarray}
{\cal L}_{yk}(\hat \phi,\hat \lambda) ={\cal L}_{yk}(  \phi, \lambda),
\end{eqnarray}
where $\phi$ stands for all the fields collectively and $\hat
\lambda$ is defined by 
\begin{equation}
\hat \lambda = \lambda +\tau.
\end{equation}
The form invariance implies that the quantum fields $\hat \phi$ and $ \phi$ satisfy the same 
equations of motion
with gauge parameters $\hat \lambda$ and $\lambda$ respectively. 

Now,  the BRST transformation for the Lagrangian density (\ref{yk}) is 
given by
\begin{eqnarray}
s  h_{ab} &=&   -(\nabla_a c_b +\nabla_b c_a ), \nonumber\\ 
s c^a  &= & -c_b\nabla^b c^a,\ \     \ s  \bar c^a
= b^a,\nonumber\\
 s  b^a & =&  0,\ \ s y^a =0,\ \ s y^a_\star =0.  
\end{eqnarray}
Corresponding to the above BRST invariance, there
exists a Noether current $J_\mu$
  satisfying the conservation law
\begin{equation}
\partial_\mu J^\mu =0,
\end{equation}
which has  the nilpotent BRST charge $Q_b =\int d^3x \sqrt{ -g}J^0$.
To define the physical states, the   unphysical gaugeon and graviton modes are removed by   
imposing the following two subsidiary conditions:
\begin{eqnarray}
&&Q_b|\mbox{phys}\rangle =0,\nonumber\\
&& y_\star^{a(+)}|\mbox{phys}\rangle =0,\label{kogo}
\end{eqnarray}
where the  first Kugo-Ojima type condition removes the unphysical gauge 
modes from the total Fock space   and the second   Gupta-Bleuler
type condition   removes the unphysical gaugeon modes.
The second subsidiary condition makes sense when the field $y^a_\star $ satisfies the
following free field equation
\begin{equation}
\nabla_b\nabla^b y^a_\star  =0,
\end{equation}
which is derived by using the equations of motion of the field $y^a$. Here the
d'Alembertian is defined as $\nabla_b\nabla^b =\frac{1}{\sqrt{-g}}\partial_\mu[\sqrt{-g} g^{ab} \partial_b]$. This free field equation guarantees the  well-defined decomposition of 
the field $y^a_\star$ into positive and negative frequency
parts. 
Therefore, for the second subsidiary condition  it is mandatory for $y^a_\star$  to satisfy the free equation.
However, for the Kugo-Ojima type condition based on the
conserved charge one has no such kind of limitation.
 
\section{Gaugeon formalism with BRST symmetry}
In this section, we develop the BRST symmetric  gaugeon formalism for perturbative quantum gravity
where the two subsidiary conditions obtained in the previous section get replaced by single 
Kugo-Ojima type subsidiary condition. 
With this motivation, we construct the
BRST symmetric  Yokoyama Lagrangian density as
\begin{eqnarray}
{\cal L}_{ykb}& =& {\cal L}_c +\sqrt{- g} b^a(\nabla^b h_{ab}-k \nabla_a h)  +\frac{\varepsilon}{2} (
y^a_\star + \lambda b^a )^2 
+\sqrt{ -g}\bar c^a M_{ab} c^b \nonumber\\
&+&  \sqrt{- g} \nabla^b y_\star ^a [ \nabla_a y_b+ \nabla_b y_a- 
2kg_{ab}\nabla_c y^c] + \sqrt{ -g}  K_\star^a M_{ab} K^b, \label{ykb}
\end{eqnarray} 
 where $K_\star^a$ and $K^a$ are Faddeev-Popov ghosts corresponding to
 gaugeon fields $y^a_\star$ and $y^a$.
 The BRST transformation for the above Lagrangian density  is given by
 \begin{eqnarray}
&&s  h_{ab} =   -(\nabla_a c_b +\nabla_b c_a), \nonumber\\ 
&&s c^a  = -c_b\nabla^b c^a,\    \  \ s  \bar c^a
= b^a,\nonumber\\
&&   
 s  b^a =  0, \ \ s y^a =K^a,\  \ sy^a_\star =0,   \nonumber\\
&& s K^a_\star  =y^a_\star,\ \ s K^a =0.\label{brs}
\end{eqnarray}
It is easy to check the nilpotency (i.e. $s^2 =0$) of the above BRST transformation. 
We further  recast the   Lagrangian density (\ref{ykb}) with the help of 
 the above BRST transformation, where the gauge-fixing and ghost parts are the
BRST variation of the extended gauge-fixing fermion, as follows:
\begin{eqnarray}
{\cal L}_{ykb}& =& {\cal L}_c  + s\sqrt{- g} \left[\bar c^a \left(\nabla^b h_{ab} -k\nabla_a h +\frac{\varepsilon\lambda}{2} (y_{a\star} +\lambda b_a)\right)\right.\nonumber\\
&-&\left. K_\star^a \left(
M_{ab} y^b -\frac{\varepsilon}{2} (y_{a\star} +\lambda b_a)\right)\right].
\end{eqnarray} 
Here the expression for the extended gauge-fixing fermion is given as
\begin{eqnarray}
\Psi & =& \sqrt{- g} \left[\bar c^a \left(\nabla^b h_{ab} -k\nabla_a h +\frac{\varepsilon\lambda}{2} (y_{a\star} +\lambda b_a)\right)\right.\nonumber\\
&-&\left. K_\star^a \left(
M_{ab} y^b -\frac{\varepsilon}{2} (y_{a\star} +\lambda b_a)\right)\right]. 
\end{eqnarray}
This gauge-fixing fermion gets identified with the gauge-fixing fermion given in the Eq. (\ref{gff})
for vanishing gaugeon and corresponding ghost fields. 

Now,  the Noether charge ($Q$) corresponding to the
BRST symmetry transformation Eq. (\ref{brs}) annihilates the physical states of the
total Hilbert space as follows: 
\begin{eqnarray}
 Q |\mbox{phys}\rangle =0,\label{kug}
\end{eqnarray}
which helps in defining the    physical Hilbert space of the theory.
This single subsidiary condition removes both the gaugeon modes and the unphysical
graviton (gauge) modes from the physical subspace of states, as 
the BRST operator acts  on both the gaugeon fields and the usual gauge fields.
(For example, it can be seen from
expression (\ref{brs}) that the gaugeon fields $y, y_\star, K$ and $K_\star$ form
a BRST quartet  which  appears only as zero-normed states
in the physical subspace \cite{ko}.)
Unlike  the Gupta-Bleuler type condition, this single condition (\ref{kug}) does not have any kind of limitation. 

Now, we establish the quantum gauge transformations under which the BRST invariant Yokoyama 
Lagrangian density (\ref{ykb}) 
remains form invariant. These transformations are given by 
\begin{eqnarray}
&& h_{ab} \rightarrow\hat h_{ab} =   h_{ab} -\tau( \nabla_a y_b +  \nabla_b y_a),\nonumber\\
&&y^a_\star\rightarrow \hat y^a_\star = y^a_\star - \tau b^a,\nonumber\\
&& y_a\rightarrow\hat y_a =y_a, 
\nonumber\\
&&b^a\rightarrow \hat b^a =b^a,\nonumber\\
&& \bar c^a\rightarrow \hat{\bar c}^a  =\bar c^a,\nonumber\\
&& c^a\rightarrow \hat c ^a =c^a
+\tau K^a,\nonumber\\
&& K^a_\star \rightarrow  \hat K^a_\star = K^a_\star -\tau \bar c^a,\nonumber\\
&&   K^a \rightarrow  \hat K^a = K^a, \nonumber\\
&& \lambda \rightarrow \hat \lambda = \lambda +\tau.
 \end{eqnarray}
 It is straightforward to check that these transformations commute with the
BRST transformation given in (\ref{brs}) which confirms that the BRST charge $Q$  remains
unchanged under the above quantum gauge transformations
\begin{equation}
Q\rightarrow \hat Q  = Q,
\end{equation} 
where $\hat Q $ is the transformed BRST charge under the quantum gauge transformations.
Therefore, the physical
space of states  $ {\cal V}_{phys}$ annihilated by the charge $Q$ also remains intact under
these transformations, i.e.
 \begin{equation}
\hat {\cal V}_{phys} =  {\cal V}_{phys}.
\end{equation} 
As a result the physical Hilbert space of quantum gravity ${\cal H}_{phys}=  {\cal V}_{phys}/\mbox{Im} Q$
is also invariant under both the BRST and the quantum gauge transformations.
\section{  Conclusions}
In this paper, we have studied the BRST symmetry for perturbative quantum gravity in    
 general curved 
spacetime
with a covariant gauge condition. 
Further, we have analyzed the Yokoyama gaugeon formalism  for the  theory of quantum gravity
and discussed the quantum gauge degree of freedom.  
Within the analysis, we have constructed the Yokoyama Lagrangian density for
the theory of quantum gravity by incorporating
two vector gaugeon fields. The quantum gauge transformations have also been investigated under which
the Yokoyama Lagrangian density for perturbative quantum gravity remains
form invariant with a shift in the gauge parameter. It has been noticed that there exist unphysical
modes also associated with both gaugeon and graviton fields and therefore
one needs to remove them from the physical Hilbert space.
We have removed them  by imposing two subsidiary conditions, the Kugo-Ojima type and 
Gupta-Bleuler type.
The Kugo-Ojima type subsidiary condition removes the unphysical gauge  modes and the Gupta-Bleuler type condition removes the unphysical gaugeon modes.  Moreover,  for the Gupta-Bleuler type condition we have found a certain limitation.

Further, the  BRST symmetric gaugeon formalism 
 has been developed for the gravity theory which incorporates  ghost fields also corresponding to
  gaugeon fields. The supremacy of the BRST version of the gaugeon formalism 
is that here the Yokoyama's physical subsidiary condition   of  Gupta-Bleuler type  translates into a more acceptable Kugo-Ojima type condition.
The BRST symmetric Yokoyama Lagrangian density possesses
 the  quantum gauge transformation also,  which  commutes
with the BRST symmetry of the theory. As a result, we have found that the physical state  annihilated by the BRST charge is also invariant under quantum gauge transformations.
Hence,  a physical Hilbert space of perturbative quantum gravity has been
constructed which remains invariant  under quantum gauge transformations.  
We hope that such an analysis will be helpful in developing the full quantum theory of gravity. 
It will be interesting to generalize the quantum gauge transformations by making the 
bosonic transformation parameter field-dependent which will lead to field-dependent 
Jacobian \cite{sudh}. It will also be interesting to
explore the gaugeon formulation in the framework of generalized BRST transformation
\cite{sud1}.

\end{document}